\documentclass[12pt,english]{article}
\usepackage{mathptmx}
\usepackage[T1]{fontenc}
\usepackage[latin9]{inputenc}
\usepackage{geometry}
\usepackage{babel}
\usepackage{float}
\usepackage{units}
\usepackage{bm}
\usepackage{amsmath}
\usepackage{amssymb}
\usepackage{graphicx}
\usepackage{setspace}
\usepackage[authoryear]{natbib}
\usepackage{xcolor}
\usepackage[unicode=true,
 bookmarks=true,bookmarksnumbered=false,bookmarksopen=false,
 breaklinks=false,pdfborder={0 0 1},backref=false,colorlinks=false]
 {hyperref}

\makeatletter

\providecommand{\tabularnewline}{\\}

\usepackage{bm}
\let\newcommand=\providecommand

\usepackage[format=hang,font=small,labelfont=bf]{caption}

\makeatother

\begin{document}
\begin{center}
\textbf{\Large{}Bayesian, frequentist and fiducial intervals for the difference between two binomial proportions}{\Large\par}
\par\end{center}

\begin{center}
Lizanne Raubenheimer$^{1}$\footnote{Corresponding author. E-mail: L.Raubenheimer@ru.ac.za}\\
{\small{}$^{1}$Department of Statistics, Rhodes University, Makhanda (Grahamstown), South Africa}\\
\par\end{center}
\begin{abstract}
Estimating the difference between two binomial proportions will be investigated, where Bayesian, frequentist and fiducial (BFF) methods will be considered. Three vague priors will be used, the Jeffreys prior, a divergence prior and the probability matching prior. A probability matching prior is a prior distribution under which the posterior probabilities of certain regions coincide with their coverage probabilities. Fiducial inference can be viewed as a procedure that obtains a measure on a parameter space while assuming less than what Bayesian inference does, i.e. no prior. Fisher introduced the idea of fiducial probability and fiducial inference. In some cases the fiducial distribution is equivalent to the Jeffreys posterior. The performance of the Jeffreys prior, divergence prior and the probability matching prior will be compared to a fiducial method and other classical methods of constructing confidence intervals for the difference between two independent binomial parameters. These intervals will be compared and evaluated by looking at their coverage rates and average interval lengths. The probability matching and divergence priors perform better than the Jeffreys prior.
\end{abstract}

\section{Introduction}

Bayesian, frequentist and fiducial (BFF) methods will be considerd.
Confidence intervals for the difference between two binomial proportions
will be constructed. The well-known Wald interval and the Agresti-Caffo
interval will be considered for the frequentist methods. For the Bayesian
methods, three priors will be considered, the Jeffreys' prior, a matching
prior and a divergence prior. A fiducial quantity will also be considered,
where a fiducial confidence interval for the difference between two
proportions will be constructed.

We assume that $X_{1},X_{2}$ are independent binomial random variables
with $X_{i}\sim Bin\left(n_{i},p_{i}\right)$ for $i=1,2.$ And we
are interested in CIs for $\delta=p_{1}-p_{2}$, the difference between
two binomial proportions. 

\section{Different Intervals}

\subsection{Frequentist }

The well-known Wald interval will be considered,
\begin{center}
${\displaystyle \left(\hat{p}_{1}-\hat{p}_{2}\right)}\pm z_{\nicefrac{\alpha}{2}}\sqrt{\frac{\hat{p}_{1}\left(1-\hat{p}_{1}\right)}{n_{1}}+\frac{\hat{p}_{2}\left(1-\hat{p}_{2}\right)}{n_{2}}}.$
\par\end{center}

As well as the Agresti-Caffo interval,
\begin{center}
${\displaystyle \left(\hat{p}_{1}^{*}-\hat{p}_{2}^{*}\right)}\pm z_{\nicefrac{\alpha}{2}}\sqrt{\frac{\hat{p}_{1}^{*}\left(1-\hat{p}_{1}^{*}\right)}{n_{1}^{*}}+\frac{\hat{p}_{2}^{*}\left(1-\hat{p}_{2}^{*}\right)}{n_{2}^{*}}}.$
\par\end{center}

where $n_{i}^{*}=n_{i}+2$, $\hat{p}_{i}^{*}=\nicefrac{\left(x_{i}+1\right)}{n_{i}^{*}}$,
and $z_{\nicefrac{\alpha}{2}}$ is the upper $\nicefrac{\alpha}{2}$
quantile of the standard normal. These intervals were considered in
Roths \& Tebbs (2006).

\subsection{Fiducial }

The concept of fiducial probability was introduced by Fisher (1930).
The following are needed for the fiducial approach:
\begin{itemize}
\item A sufficient statistic for the parameter in question.
\item A pivot, function of both sufficient statistic and true value of the
parameter.
\item The fiducial argument - from the distribution of the pivot, a distribution
for the parameter can be derived on the basis of the sampled sufficient
statistic.
\end{itemize}
Let $X\sim Bin\left(n,p\right)$ and let $B_{a,b}$ denote the beta
RV with parameters $a$ and $b$. For an observed value $x$ of $X$,
$P\left(X\geq x\left|n,p\right.\right)=P\left(B_{x,n-x+1}\leq p\right)$
and $P\left(X\leq x\left|n,p\right.\right)=P\left(B_{x+1,n-x}\geq p\right)$. 

From this we can see that there is a pair of fiducial distributions
for $p$: 
\begin{itemize}
\item $B_{x,n-x+1}$ 
\item $B_{x+1,n-x}$ 
\end{itemize}
As stated in Krishnamoorthy \& Zhang (2015), instead of having two fiducial variables,
a random quantity that is between $B_{x,n-x+1}$ and $B_{x+1,n-x}$
can be used as a single approximate fiducial variable for $p$. From
Cai (2005), a simple choice is $B_{x+0.5,n-x+0.5}$. 

We are interested in a fiducial quantity for $p_{1}-p_{2}$. Let $x_{1}$
be an observed value of $X_{1}$ and $x_{2}$ be an observed value
of $X_{2}$. 

The fiducial quantity for $p_{i}$ is given by $B_{x_{i}+0.5,n_{i}-x_{i}+0.5}$
for $i=1,2$.

The fiducial quantity for the difference $\delta=p_{1}-p_{2}$ is
given 
\begin{align*}
Q_{\delta} & =B_{x_{1}+0.5,n_{1}-x_{1}+0.5}-B_{x_{2}+0.5,n_{2}-x_{2}+0.5.}
\end{align*}
The $1-\alpha$ fiducial CI is given by
\[
\left(Q_{\delta,\frac{\alpha}{2}},Q_{\delta,1-\frac{\alpha}{2}}\right).
\]

Note: This fiducial CI gives the same result as a Bayesian CI when
using the Jeffreys prior.

\subsection{Bayesian intervals}
\begin{itemize}
\item \textbf{Jeffreys' Prior:}

\begin{align*}
\pi_{J}\left(p_{1},p_{2}\right) & \propto\prod_{i=1}^{2}p_{i}^{-\frac{1}{2}}\left(1-p_{i}\right)^{-\frac{1}{2}}.
\end{align*}

The resulting joint posterior distribution will then be

\begin{align*}
\pi_{J}\left(p_{1},p_{2}\left|x_{1},x_{2}\right.\right) & =\frac{1}{B\left(x_{1}+\frac{1}{2},n_{1}-x_{1}+\frac{1}{2}\right)}p_{1}^{x_{1}-\frac{1}{2}}\left(1-p_{1}\right)^{n_{1}-x_{1}-\frac{1}{2}}\\
 & \times\frac{1}{B\left(x_{2}+\frac{1}{2},n_{2}-x_{2}+\frac{1}{2}\right)}p_{2}^{x_{2}-\frac{1}{2}}\left(1-p_{2}\right)^{n_{2}-x_{2}-\frac{1}{2}}.
\end{align*}

\item \textbf{Divergence Prior:}

\begin{align*}
\pi_{D}\left(p_{1},p_{2}\right) & \propto\prod_{i=1}^{2}p_{i}^{-\frac{1}{4}}\left(1-p_{i}\right)^{-\frac{1}{4}}.
\end{align*}

The resulting joint posterior distribution will then be

\begin{align*}
\pi_{D}\left(p_{1},p_{2}\left|x_{1},x_{2}\right.\right) & =\frac{1}{B\left(x_{1}+\frac{3}{4},n_{1}-x_{1}+\frac{3}{4}\right)}p_{1}^{x_{1}-\frac{1}{4}}\left(1-p_{1}\right)^{n_{1}-x_{1}-\frac{1}{4}}\\
 & \times\frac{1}{B\left(x_{2}+\frac{3}{4},n_{2}-x_{2}+\frac{3}{4}\right)}p_{2}^{x_{2}-\frac{1}{4}}\left(1-p_{2}\right)^{n_{2}-x_{2}-\frac{1}{4}}.
\end{align*}

\item \textbf{Probability Matching Prior:}

Datta \& Ghosh (1995) derived the differential equation which a prior must
satisfy if the posterior probability of a one sided credibility interval
for a parametric function and its frequentist probability agree up
to a certain order. They proved that the agreement between the posterior
probability and the frequentist probability holds if and only if $\sum\limits _{i=1}^{k}\frac{\partial}{\partial p_{i}}\left\{ \eta_{i}\left(\underline{p}\right)\pi\left(\underline{p}\right)\right\} =0.$ 

\textbf{Theorem 1} Assume that $X_{1},X_{2}$ are independent Binomial
random variables with $X_{i}\sim Bin\left(n_{i},p_{i}\right)$ for
$i=1,2.$ The probability matching prior for $\delta=p_{1}-p_{2}$,
the difference between two Binomial proportions, is given by 

\begin{center}
$\pi{}_{M}\left(p_{1},p_{2}\right)\propto\left\{ \sum\limits _{i=1}^{2}p_{i}\left(1-p_{i}\right)\right\} ^{\frac{1}{2}}\prod\limits _{i=1}^{2}p_{i}^{-1}\left(1-p_{i}\right)^{-1}.$
\par\end{center}

The resulting joint posterior distribution will then be

\begin{center}
$\pi{}_{M}\left(p_{1},p_{2}\left|x_{1},x_{2}\right.\right)\propto\left\{ \sum\limits _{i=1}^{2}p_{i}\left(1-p_{i}\right)\right\} ^{\frac{1}{2}}\prod\limits _{i=1}^{2}p_{i}^{x_{i}-1}\left(1-p_{i}\right)^{n_{i}-x_{i}-1}.$
\par\end{center}

It was shown in Raubenheimer \& Van der Merwe (2011) that this posterior distribution is
proper.
\end{itemize}

\section{Simulation Study}

In this section a simulation study will be done, where coverage rates
and average interval lengths for $p_{1}-p_{2}$ will be obtained.
Raubenheimer \& Van der Merwe (2011) compared the performance of the Jeffreys and probability
matching priors with that of the frequentist results.
\begin{center}
\textbf{}
\begin{table}[H]
\caption{Exact coverage rates (CR) and average interval lengths (LE), the nominal
level is 0.95.}

\begin{centering}
\begin{tabular}{|c|c|c|c|c|c|}
\hline 
 & WAL & AGC & $\pi_{J}$/ FID & $\pi_{D}$ & $\pi_{M}$\tabularnewline
\hline 
\multicolumn{6}{|c|}{$n_{1}=n_{2}=10$}\tabularnewline
\hline 
\textcolor{gray}{CR} & 0.917 & 0.963 & 0.945 & 0.954 & 0.948\tabularnewline
\textcolor{gray}{LE} & 0.659 & 0.680 & 0.650 & 0.667 & 0.649\tabularnewline
\hline 
\multicolumn{6}{|c|}{$n_{1}=n_{2}=20$}\tabularnewline
\hline 
\textcolor{gray}{CR} & 0.931 & 0.958 & 0.945 & \textcolor{blue}{0.949} & \textcolor{blue}{0.949}\tabularnewline
\textcolor{gray}{LE} & 0.489 & 0.494 & 0.482 & 0.497 & \textcolor{blue}{0.481}\tabularnewline
\hline 
\end{tabular}
\par\end{centering}
\end{table}
\par\end{center}

\begin{table}[H]

\caption{Exact coverage rates (CR) and average interval lengths (LE), the nominal
level is 0.95, $n_{1}=n_{2}=10$.}

\begin{centering}
\begin{tabular}{|c|c|c|c|c|c|c|c|}
\hline 
$p_{1}$ & $p_{2}$ &  & WAL & AGC & $\pi_{J}$/ FID & $\pi_{D}$ & $\pi_{M}$\tabularnewline
\hline 
\textcolor{gray}{0.1} & \textcolor{gray}{0.1} & \textcolor{gray}{CR} & \textcolor{blue}{0.950} & 0.991 & 0.967 & 0.991 & 0.988\tabularnewline
 &  & \textcolor{gray}{LE} & \textcolor{blue}{0.456} & 0.578 & 0.542 & 0.559 & 0.568\tabularnewline
\hline 
\textcolor{gray}{0.1} & \textcolor{gray}{0.7} & \textcolor{gray}{CR} & 0.915 & 0.945 & 0.945 & \textcolor{blue}{0.947} & 0.937\tabularnewline
 &  & \textcolor{gray}{LE} & 0.619 & 0.656 & \textcolor{blue}{0.617} & 0.619 & \textcolor{blue}{0.617}\tabularnewline
\hline 
\textcolor{gray}{0.3} & \textcolor{gray}{0.3} & \textcolor{gray}{CR} & 0.905 & 0.963 & 0.937 & \textcolor{blue}{0.962} & 0.968\tabularnewline
 &  & \textcolor{gray}{LE} & 0.756 & 0.727 & 0.717 & 0.708 & \textcolor{blue}{0.699}\tabularnewline
\hline 
\textcolor{gray}{0.3} & \textcolor{gray}{0.7} & \textcolor{gray}{CR} & 0.932 & 0.955 & 0.935 & \textcolor{blue}{0.948} & 0.943\tabularnewline
 &  & \textcolor{gray}{LE} & 0.754 & 0.727 & 0.707 & 0.698 & \textcolor{blue}{0.690}\tabularnewline
\hline 
\textcolor{gray}{0.5} & \textcolor{gray}{0.5} & \textcolor{gray}{CR} & 0.912 & 0.958 & \textcolor{blue}{0.956} & \textcolor{blue}{0.956} & 0.960\tabularnewline
 &  & \textcolor{gray}{LE} & 0.830 & 0.771 & 0.763 & 0.751 & \textcolor{blue}{0.741}\tabularnewline
\hline 
\end{tabular}
\par\end{centering}
\end{table}

\begin{center}
\textbf{}
\begin{table}[H]

\textbf{\caption{Exact coverage rates (CR) and average interval lengths (LE), the nominal
level is 0.95, $n_{1}=n_{2}=20$.}
}
\begin{centering}
\begin{tabular}{|c|c|c|c|c|c|c|c|}
\hline 
$p_{1}$ & $p_{2}$ &  & WAL & AGC & $\pi_{J}$/ FID & $\pi_{D}$ & $\pi_{M}$\tabularnewline
\hline 
\textcolor{gray}{0.1} & \textcolor{gray}{0.1} & \textcolor{gray}{CR} & \textcolor{blue}{0.960} & 0.988 & 0.942 & 0.961 & 0.963\tabularnewline
 &  & \textcolor{gray}{LE} & \textcolor{blue}{0.352} & 0.396 & 0.378 & 0.387 & 0.394\tabularnewline
\hline 
\textcolor{gray}{0.1} & \textcolor{gray}{0.7} & \textcolor{gray}{CR} & 0.913 & 0.955 & \textcolor{blue}{0.948} & \textcolor{blue}{0.953} & 0.941\tabularnewline
 &  & \textcolor{gray}{LE} & 0.464 & 0.473 & \textcolor{blue}{0.456} & \textcolor{blue}{0.456} & 0.459\tabularnewline
\hline 
\textcolor{gray}{0.3} & \textcolor{gray}{0.3} & \textcolor{gray}{CR} & 0.931 & \textcolor{blue}{0.950} & 0.940 & \textcolor{blue}{0.950} & 0.957\tabularnewline
 &  & \textcolor{gray}{LE} & 0.552 & 0.538 & 0.534 & 0.531 & \textcolor{blue}{0.528}\tabularnewline
\hline 
\textcolor{gray}{0.3} & \textcolor{gray}{0.7} & \textcolor{gray}{CR} & 0.928 & \textcolor{blue}{0.944} & 0.942 & \textcolor{blue}{0.944} & \textcolor{blue}{0.956}\tabularnewline
 &  & \textcolor{gray}{LE} & 0.552 & 0.538 & 0.499 & 0.528 & 0.522\tabularnewline
\hline 
\textcolor{gray}{0.5} & \textcolor{gray}{0.5} & \textcolor{gray}{CR} & 0.919 & 0.957 & \textcolor{blue}{0.950} & 0.957 & 0.958\tabularnewline
 &  & \textcolor{gray}{LE} & 0.604 & 0.578 & 0.576 & 0.571 & \textcolor{blue}{0.564}\tabularnewline
\hline 
\end{tabular}
\par\end{centering}
\end{table}
\par\end{center}

\section{Illustrative Example}

Consider an example from Ornaghi et al. (1999). The goal of this experiment
was to assess if male and female insects transmit the Mal de Rio Cuarto
virus to susceptible maize plants at similar rates.

Assume that, at a specific stage, the researchers want to estimate
the difference $p_{1}-p_{2}$, where $p_{1}$ is equal to the proportion
infected plants for male insects and $p_{2}$ is the proportion infected
plants for female insects. Stage 1 will be considered here, where
$\hat{p}_{1}=\nicefrac{9}{29}$ and $\hat{p}_{2}=\nicefrac{5}{31}$.

\begin{table}[H]

\caption{95\% Credibility Intervals for the Difference in Disease Transmission
Probabilities among Male and Female insects.}

\begin{centering}
\begin{tabular}{|c|c|c|c|c|c|c|}
\hline 
 & \textbf{Interval} & WAL & AGC & $\pi_{J}$/ FID & $\pi_{D}$ & $\pi_{M}$\tabularnewline
\hline 
\hline 
\textbf{Stage} & Lower limit & -0.063 & -0.070 & -0.060 & -0.063 & -0.062\tabularnewline
\cline{2-7} \cline{3-7} \cline{4-7} \cline{5-7} \cline{6-7} \cline{7-7} 
\textbf{1} & Upper limit & 0.631 & 0.351 & 0.352 & 0.348 & 0.345\tabularnewline
\cline{2-7} \cline{3-7} \cline{4-7} \cline{5-7} \cline{6-7} \cline{7-7} 
 & Length & 0.425 & 0.421 & 0.412 & 0.411 & 0.406\tabularnewline
\hline 
\end{tabular}
\par\end{centering}
\end{table}

\section{Conclusion}

In this paper the probability matching prior for the difference between
two Binomial rates were derived. Limited simulation studies have shown
that the probability matching prior achieves its sample frequentist
coverage results somewhat better than in the case of the Jeffreys'
prior. The matching prior and the divergence prior yielded similar
results. The fiducial CI is the same as the Bayesian CI when using
the Jeffreys' prior.

\section*{References}
\begin{list}{}{}
\item Cai, T. (2005). One-sided confidence intervals in discrete distributions. \emph{Journal of Statistical Planning
and Inference}, 131, 63-88.
\item Datta, G. S. \&  Ghosh, J. K. (1995). On priors providing frequentist validity for Bayesian inference.
\emph{Biometrika}, 82(1), 37-45.
\item Fisher, R. A. (1930). Inverse probability.\emph{ Proceedings of the Cambridge Philosophical Society}, 26, 528-535.
\item Krishnamoorthy, K. \&  Zhang, D. (2015). Approximate and fiducial confidence intervals for the difference
between two binomial proportions. \emph{Communications in Statistics - Theory and Methods}, 44, 1745-1759.
\item Ornaghi, J. A., March, G. J., Boito, G. T., Marinelli, A., Beviacqua, J. E., Giuggia, J., \&  Lenardon, S. L.
(1999). Infectivity in natural populations of Delphacodes kuscheli vector of "Mal Rio Cuarto" virus.
\emph{Maydica}, 44, 219-223.
\item Raubenheimer, L. \&  Van der Merwe, A. J. (2011). Bayesian estimation of functions of binomial rates.
\emph{South African Statistical Journal}, 45(1), 41-64.
\item Roths, S. A. \&  Tebbs, J. M. (2006). Revisiting Beal's confidence intervals for the difference of two
binomial proportions. \emph{Communications in Statistics - Theory and Methods}, 35(9), 1593-1609.
\end{list}{}{} 

\end{document}